\documentclass[%
 reprint,
 amsmath,amssymb,
 aps,
]{revtex4-2}

\usepackage{graphicx}% Include figure files
\usepackage{dcolumn}% Align table columns on decimal point
\usepackage{bm}% bold math
\usepackage[dvipsnames]{xcolor}
\usepackage{dcolumn}% Align table columns on decimal point
\usepackage{bm}% bold math environment with \bm{}
\usepackage{dsfont}  % To use \mathds{} correctly.
\usepackage{bbold}  % To use \mathbb{} correctly.
\usepackage{empheq}  % To get a refined structure on equations like big bracket for a set of equations.
\usepackage{amsmath}
\usepackage{ulem}
\usepackage{lipsum}% http://ctan.org/pkg/lipsum
\usepackage{comment}
\begin{document}

\preprint{APS/123-QED}

%\title{Spectrally Multimode Squeezed State Generation at Telecommunication Wavelengths}
\title{Multimode Squeezed State for Reconfigurable \\ Quantum Networks at Telecommunication Wavelengths}

\author{Victor Roman-Rodriguez$^{1,2}$}\email{victor.roman@icfo.eu}\altaffiliation[Currently with ]{ICFO - Instituto de Ciencias Fotonicas, The
Barcelona Institute of Science and Technology, 08860
Castelldefels, Barcelona, Spain}\author{David Fainsin$^1$}\author{Guilherme L. Zanin$^1$, Nicolas Treps$^1$, Eleni Diamanti$^2$, and Valentina Parigi$^{1}$}
\affiliation{$^1$Laboratoire Kastler Brossel, Sorbonne Universit\'e, ENS-Université PSL, CNRS, Coll\`ege de France, 4 place Jussieu, Paris F-75252, France}
\affiliation{$^2$Sorbonne Universit\'e, LIP6, CNRS, 4 place Jussieu, 75005 Paris, France}

\date{\today}% It is always \today, today,
\begin{abstract} 
Continuous variable encoding of quantum information requires the deterministic generation of highly correlated quantum states of light in the form of quantum networks, which, in turn, necessitates the controlled generation of a large number of squeezed modes. In this work, we present an experimental  source of multimode squeezed states of light at telecommunication wavelengths. Generation at such wavelengths is especially important as it can enable quantum information processing, communication, and sensing beyond the laboratory scale.
We use a single-pass spontaneous parametric down-conversion process in a non-linear waveguide pumped with the second harmonic of a femtosecond laser. Our measurements reveal significant squeezing in more than 21 frequency modes, with a maximum squeezing value exceeding 2.5 dB. We demonstrate multiparty entanglement by measuring the state's covariance matrix. Finally, we show the source reconfigurability by preparing few-node cluster states and measure their nullifier squeezing level.
These results pave the way for a scalable implementation of continuous variable quantum information protocols at telecommunication wavelengths, particularly for multiparty, entanglement-based quantum communications. Moreover, the source is compatible with additional pulse-by-pulse multiplexing, which can be utilized to construct the necessary three-dimensional entangled structures for quantum computing protocols.
\end{abstract}

%\begin{abstract} 
%Continuous variable encoding of quantum information requires the deterministic generation of highly correlated quantum states of light in the form of quantum networks, which, in turn, necessitates the controlled generation of a large number of squeezed modes. In this work, we present an experimental demonstration of a source that generates multimode squeezed states of light at telecommunication wavelengths. Generation at such wavelengths is especially important, as it can enable quantum information, communication, and sensing beyond the laboratory scale.
%We achieve this using a single-pass spontaneous parametric down-conversion process in a non-linear waveguide pumped with a femtosecond laser. Our measurements reveal significant squeezing in more than 21 frequency modes, with a maximum squeezing value exceeding 2.5 dB. We demonstrate multiparty entanglement by measuring the state's covariance matrix. Finally, we show the source reconfigurability by preparing few-node cluster states and measure their nullifier squeezing level.
%This result paves the way for a scalable implementation of continuous variable quantum information protocols at telecommunication wavelengths, particularly for multiparty and entanglement-based quantum communications. Moreover, the source is compatible with additional pulse-by-pulse multiplexing, which can be utilized to construct the necessary three-dimensional entangled structures for quantum computing protocols.
%\end{abstract}

\maketitle

%\tableofcontents

%\section{Introduction}

Continuous variable (CV) encoding of quantum information requires the generation of multimode quantum states of light with tailored spectral, spatial and temporal mode properties \cite{SPOPO,OpticaPaderborn}.  In particular, measurement-based protocols rely on the possibility of deterministically generating large multimode entangled states, where entanglement is established between amplitude and phase quadratures of different light modes \cite{Chen14,Cai17,Yokoyama13,Asavanant19,Larsen19}, forming what can be described as quantum networks. This in turn requires the generation of a large number of squeezed modes to begin with. 

The use of second-order nonlinear waveguides is currently largely explored to generate these tailored squeezed modes:  single-mode over a large bandwidth  \cite{Inoue23,Kashiwazaki23, Nehra22} or temporally multiplexed \cite{Tomoda23}. Waveguides are best suited for compact sources compared to cavity-based structures.% and both spectrally and temporally multiplexed \cite{Koadou22}.

Moreover, particular effort has been devoted to generating squeezed sources at telecommunication wavelengths. Single-mode squeezed state \cite{Ast13,Gehring15}, on-chip few-mode squeezed states \cite{Mondain19,Lenzini18} and micro-comb structures have been shown \cite{Yang2021}. At these wavelengths optical communication technologies can be exploited \cite{Inoue23}, enabling quantum state transmission, and hence quantum communication, through fibers over long distances \cite{Suleiman2022} as well as quantum sensing \cite{Domeneguetti23}. In particular, spectral clusters can be directly used for frequency-multiplexed  QKD \cite{Kovalenko21} and entangled-based multiparty quantum communication protocols \cite{Murta20}.
%Recent results on single-mode squeezing generation in nonlinear waveguides have been reported in a regime where optical communication technologies can be exploited \cite{Inoue23}, where it is possible to have quantum state transmission through fibers over long distances \cite{Suleiman2022} and sensing \cite{Domeneguetti23}.

%Compact solutions for scalable entangled states are particularly required in quantum computing. 
%Regarding the scalability of quantum networks, temporal multiplexing has already enabled the generation of the largest CV-cluster states \cite{Yokoyama13, Asavanant19, Larsen19}.  A photonic fault-tolerant quantum computer requires three-dimensional structures \cite{Bourassa2021}, which can be realized through two-dimensional structures, compatible with the source of this work, multiplexed in time.

%Full versatility in the entanglement structure can be reached via fast electro-optical switching \cite{Madsen2022}. 
%It is in any case challenging to not add losses in mixing squeezed modes at different times, which is particularly detrimental in quantum computing.% moreover the spatial and spectral structure of the squeezed modes should be fully controlled in the mixing process and at the measurement stage. Losses and mode-mismatching reduce squeezing  (so entanglement) and the purity of the multiplexed entangled state, this is particularly detrimental in quantum computing. 

Scalable quantum networks, in the form of cluster states, are particularly required for quantum computing.  Temporal multiplexing has enabled the generation of the largest CV-cluster states to date \cite{Yokoyama13, Asavanant19, Larsen19}.  A photonic fault-tolerant quantum computer \cite{Bourassa2021} requires three-dimensional structures, along with node-selective non-Gaussian operations. While full versatility in the entanglement structure can be reached via fast electro-optical switching \cite{Madsen2022}, CV encoding also requires the ability to fully match the mode-structure in coherent (homodyne) measurement \cite{Cai17}. 

In this work we pursue scalability in the near infrared C-Band, by demonstrating a resource that is intrinsically spectrally multimode, in a regime where the full multimode structure is accessed and manipulated to form reconfigurable  2-dimensional CV networks \cite{Cai17}. Moreover such a structure is compatible with node-selective  non-Gaussian operations \cite{Ra20}, and with temporal multiplexing, as it was recently demonstrated in a single-pass configuration \cite{Kouadou2022},  thus allowing  for 3-dimensional structures \cite{Kouadou2022, Bourassa2021}. 

%Here, we demonstrate the generation of a vacuum squeezed field in the near infrared C-Band that is intrinsically spectrally multimode and that can be fully exploited in quantum protocols. The system cannot be reduced to a single squeezer on a specific spectral mode, but scales up as many squeezers on different orthogonal modes, that can be fully accessed via mode-selective homodyne detection \cite{Cai17}. In this way, the multimode state can be directly used to produce reconfigurable quantum networks in the spectral domain. Moreover, the system is compatible with  temporal multiplexing as it was recently demonstrated in a single-pass configuration \cite{Kouadou2022}.  

Here, we not only characterize the multimode spectral structure, but we also independently check entanglement correlations between different individual frequency bands. This is done by measuring the full quadrature covariance matrix and applying a Positive-Partial-Transpose (PPT) criterion. Finally, we show the source reconfigurability by generating quantum networks of different topologies and charaterizing their quality via the measurement of the so-called nullifier operators. 

%Here we demonstrate the generation of a vacuum squeezed field in the near infrared C-Band that is intrinsically spectrally multimode, {\it{i.e.}}, a system that cannot be reduced to a single squeezer acting on a specific spectral mode, but that involves many squeezers acting on different (orthogonal) spectral modes.  This implies that the many spectral modes can be directly shaped, via linear optics transformations, into entangled networks of the same number of nodes without mixing them with extra vacuum field states, thus not degrading the squeezing and/or entanglement correlations. 
%The networks built in this way can be easily tailored in our setup via mode-selective homodyne detection \cite{Cai17}, and their quality can be characterized via the squeezing level of their so-called nullifier operators. We can also independently check entanglement correlations between different individual frequency bands, by measuring the full quadrature covariance matrix for 8 different bands. A  Positive-Partial-Transpose (PPT) criterion can be used as an indication of entanglement over all the possible band bipartitions.

%Both intra-band entanglement and cluster structures can be used for frequency-multiplexed  QKD \cite{Kovalenko21} and entangled-based multiparty quantum communication protocols \cite{Murta20}.mode-selective non-Gaussian operations \cite{Ra20},  can be added to go towards computing protocols.

\begin{figure*}
	\begin{center}
		\includegraphics[width=1\textwidth]{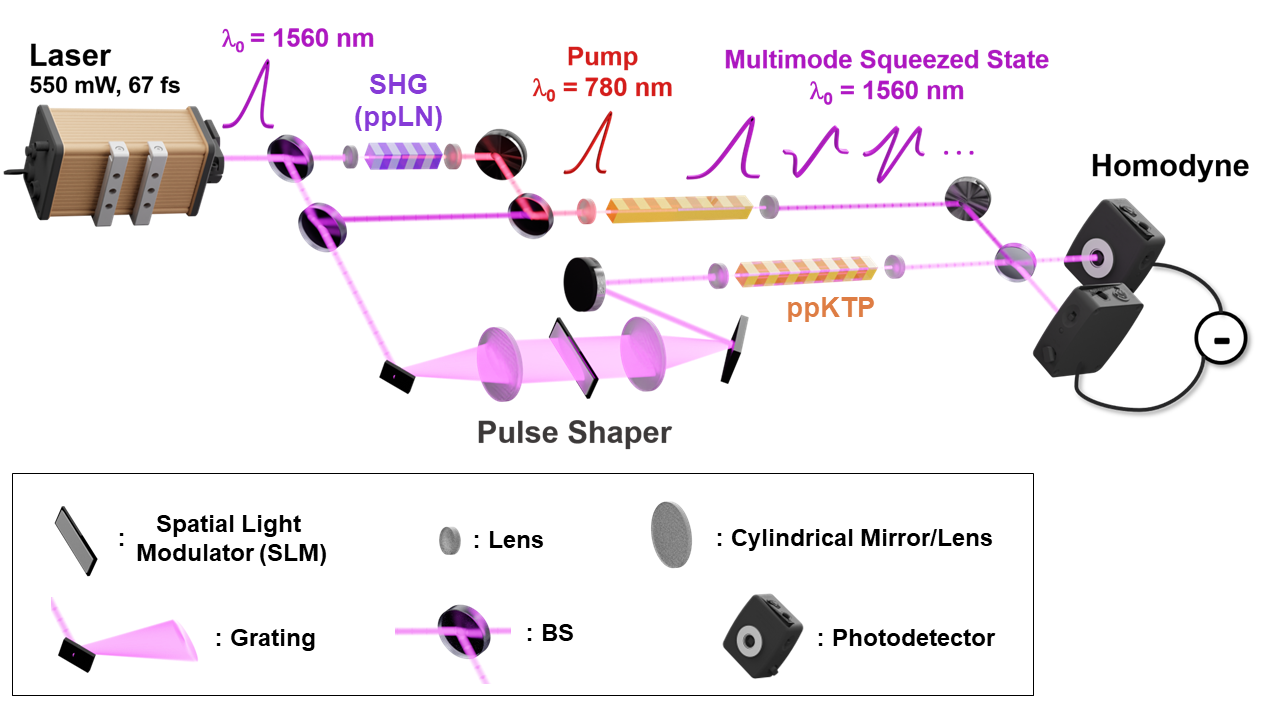}
	\end{center}
	\caption{Experimental scheme for the generation of the multimode states. A telecom wideband pulsed laser is up converted to its second harmonic with a ppLN crystal and then coupled to a nonlinear ppKTP waveguide. Type 0 SPDC interaction generates independent squeezing in a set of frequency modes. Each mode is addressed individually with homodyne detection, where the local oscillator is shaped spectrally with the use of a pulse shaper. For more details see text.}\label{fig:exp_scheme}
\end{figure*}

In our experiment we work with a type-0 waveguide in the degenerate case. The Hamiltonian describing the output field of this interaction can be written as \cite{Roslund14,VRR21}:
\begin{equation}
    \hat{H} = \sum_{k=1}^N\lambda_k\left(\hat{A}_k^\dag\right)^2 + \mathrm{h.c.}
    \label{H_AA}
\end{equation}
where $\{\lambda_k\}$ are called the Schmidt coefficients and the operators $\{\hat{A}_k\}$ are: 
\begin{equation}
    \hat{A}_k = \int{h_k(\omega)\hat{a}(\omega)\mathrm{d\omega}}
    \label{A}
\end{equation}
with $\hat{a}(\omega)$ the annihilation operator of a photon in the plane-wave basis with angular frequency $\omega$. The operator $\hat{A}_k$ is therefore an annihilation operator of photons in the frequency mode $h_k(\omega)$.

Evolution under the Hamiltonian of Eq.(\ref{H_AA}) produces a multimode field with $N$ independent squeezed states for each of the frequency modes $\{h_k\}$, as defined in Eq.(\ref{A}). Any cluster state with up to $N$ modes can then be constructed from this multimode field by an appropriate unitary transformation \cite{Menicucci11, Ferrini15}.  The specific unitary transformation can be performed by the use of passive optics or equivalently by the projection of the modes in the adequate basis via homodyne detection \cite{SPOPO,Cai17}.

\begin{figure*}
\centering
		\includegraphics[width=1\textwidth]{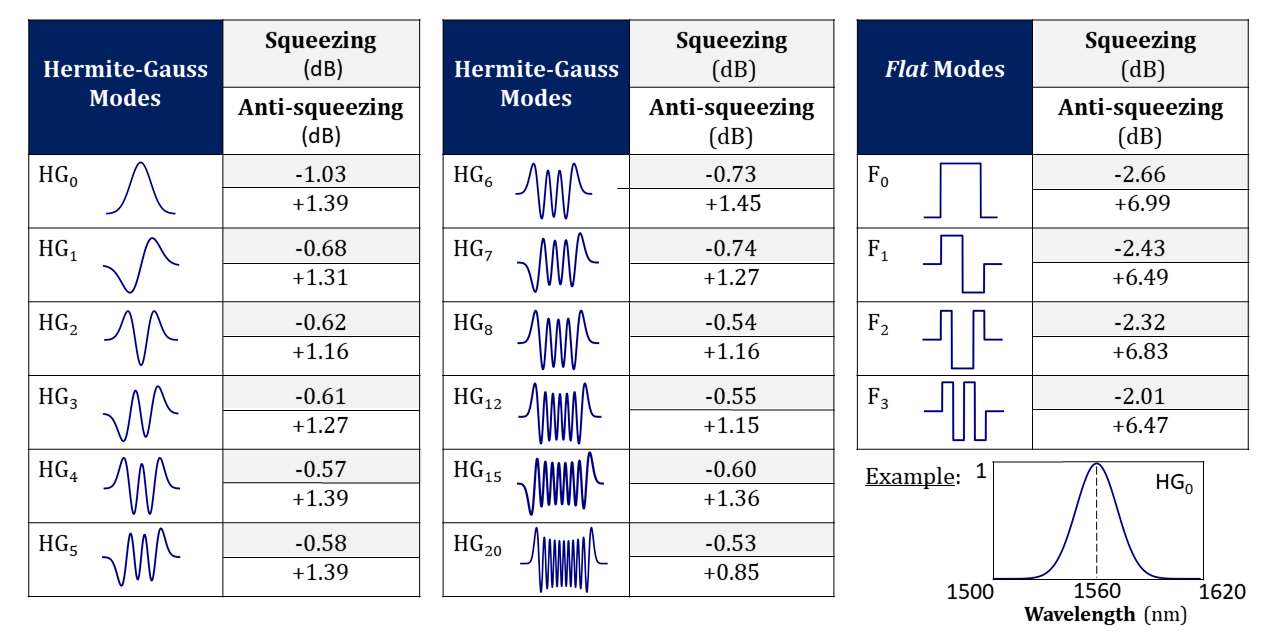}
	\caption{Experimental squeezing and anti-squeezing values for the Hermite-Gauss modes (left) and the flat modes defined in the text (right). An example of the HG$_0$ mode is also shown, indicating the spectral range common to all the modes.}\label{fig:SqzCurves}
\end{figure*}

The experimental setup for the generation of arbitrary $N$-mode cluster states is depicted schematically in Fig.~\ref{fig:exp_scheme}. A broadband fiber femtosecond laser (bandwidth $\sim$ 55 nm, pulse width $\sim$ 57 fs, repetition rate 100 MHz, power $\sim$ 550 mW, centered at 1560 nm), is partly directed to a periodically poled Lithium Niobate (ppLN) crystal, engineered to produce the second harmonic frequency. The ppLN crystal provides light with a bandwidth of $\sim$ 2 nm, centered at 780 nm. This field is then coupled to a single(spatial)-mode, rectangular, nonlinear, periodically poled Potassium Titanyl Phosphate (ppKTP) waveguide \footnote{The waveguides were purchased from the company AdvR.}, which down converts the second harmonic field to the C-band telecom wavelengths and generates multimode squeezed states according to the Hamiltonian in Eq.(\ref{H_AA}).

On the other side, a large fraction of the original power from the laser is sent to a pulse-shaper in order to generate a spectrally configurable Local Oscillator (LO). The signal field from the waveguide and the LO are mixed and directed to separate photodiodes, whose electrical outputs are subtracted to obtain the homodyne signal.

We implemented the pulse-shaper by diffracting the wavelength components of our input field using a grating, and directing them to a Spatial Light Modulator (SLM), where the light is reflected back and recombined in a similar grating (so called 4f configuration)\cite{monmayrant}. In the SLM screen, each pixel, and hence, each frequency component of the field, can be addressed individually, resulting in a  spectrally shaped pulse at the output. %We used a home-made interface in Python to control the different masks applied to the pulse shaper.

The LO is also coupled to another ppKTP waveguide (identical in dimensions) before it mixes with the quantum signal. This is done to spatially match the LO and the signal, effectively decreasing losses in the detection.

We perfomed numerical simulations to predict the properties of the independent squeezed modes following \cite{VRR21}. Details about these numerical simulations and about the characterization of our waveguides can be found in the Supplemental Material \cite{SM}.

%\section{Multimode Squeezed State Generation}
Fig.~\ref{fig:SqzCurves} summarizes the squeezing values, in dB, obtained in each of the measured modes, where the shot noise stands as reference (0 dB). The signal's quadrature noise was measured with a spectrum analyzer, as a function of the relative phase between the signal and the LO, for different LO spectral shapes. %The relative phase is modulated in time thanks to a piezoelectric mirror in the LO optical path.
%\footnote{This is why in Fig.(\ref{fig:SqzCurves}) the x-axis is time.}. 
%The spectral shapes followed the expected family of Hermite-Gauss modes for the squeezed eigenstates in the supermode basis.
We first projected the output states into the expected family of Hermite-Gauss (HG) modes, where the spectral width of the HG$_{0}$ mode in amplitude was 45 nm.

We show in Fig.~\ref{fig:SqzCurves} the measured squeezing and antisqueezing values of the first 21 HG modes. They are the minimal and maximal values of the quadrature noise measured via averaging on scans of several phase-periods from the spectrum analyzer. 
%A reduction of the homodyne noise signal under the shot noise indicates therefore the presence of squeezing, and the amount of reduction is the squeezing level. 
%Complementarily, the increase in noise in the conjugate quadrature with respect to the shot noise is the level of antisqueezing. 

The asymmetry between squeezing and antisqueezing levels is attributed to experimental optical losses, where the main loss source is the non-optimal spatial mode-matching between the signal and LO optical modes. In our case, this can be due to inhomogeneities in the two waveguide structures that are known to appear when the waveguide is long ($\sim$ cm scale). 

We characterize these losses by measuring the visibility of fringes between the LO and a small fraction of telecom light coupled into the waveguides, that resulted to be about 77\% in our setup. The associated mode-matching efficiency scales quadratically with the measured visibility. 

%In order to match the spatial modes of LO and signal, we inserted, in the LO path, a waveguide identical to the one used for the generation of the signal. 
%The overall visibility in such a configuration, which is the one of the measurements in Fig.~\ref{fig:SqzCurves},  reaches the value of 77\%. 

%In first instance, with visibilities of around 50\%, low squeezing levels could be measured for the first few supermodes. Since we can shape the LO temporaly, we concluded that the main reason for the low visibility was the spatial mismatch between the LO and the fundamental mode of the waveguide. For this reason, we added another KTP waveguide with the same dimensions into the LO optical path, so that the LO's spatial profile is naturally matched to the quantum signal.
%Thanks to this, we increased the visibility up to 

%We explain the  non-ideal visibility by residual differences between the spatial modes of the signal and the LO. This can be due to inhomogeneities in the two waveguide structures that are known to appear when the waveguide is long ($\sim$ cm scale). 

%to average impurities inside the waveguides, that are well known to appear when the poling region is long. 
%The use of shorter waveguides could therefore be a practical solution for the future setup improvement, although in that case the interaction strength would be reduced, producing lower squeezing levels. The effect of having a shorter waveguide on the number and shape of the supermodes can be predicted in advance \cite{VRR21}.
Another effect in our setup that contributes to equivalent losses is the limitation of the LO bandwidth at the wings of some frequency modes due to the limited physical size of a cylindrical mirror in our pulse shaper. This effect, which is enhanced for higher-order HG modes, is nevertheless not a fundamental limitation of the experiment. 

 %%%%%%%%%Fig.~\ref{fig:SqzCurves} also shows the spectral shapes set in the  LO pulse shaper to measure the corresponding squeezing values.

%%%%%%%It is also worth mentioning that due to the optical clipping, the cut HG modes in which we project our state are technically not orthogonal anymore. The dimension of the subspace spanned by the modes shown in Fig.~\ref{fig:SqzCurves} is nevertheless close to 21 HG modes (details can be found in Appendix E, where we calculate it to be around 18).
 
Additionally, we projected the states into a basis of orthogonal and flattened HG modes, that we call \textit{flat} modes. For such modes we observed larger squeezing values - with more than 2 dB of squeezing up to the 4th mode - than for the HG modes. The flat modes and their squeezing levels are shown in the right side of Fig.~\ref{fig:SqzCurves}. 

Examples of single squeezing traces, possible extra contributions to the global loss sources and details about the flat modes are discussed in the Supplemental Material \cite{SM}.%%%%%%%%%%%%%Although there is no evidence that the \textit{flat} modes are the supermodes of the system, this result implies that the HG basis shown on the left side of Fig.~\ref{fig:SqzCurves}  is not the family of supermodes, since those  should be  the most squeezed modes in the system. 

%%%%%%%%%%%%%%%%Further evidence in this direction is given by the covariance matrix measurements in the next section, showing that the spectral widths of the theoretically predicted HG modes are probably underestimated.

%We attribute this both to the optical clipping in the LO and the fact that the spectral widths of the HG modes that were theoretically predicted spectral were somewhat low (see discussion on the covariance matrix in the next section).

%Additionally, in Fig.(\ref{fig:SqzCurves}), we have depicted the spectral part (shown in red) of each Hermite-Gauss mode that cannot be shaped by our Spatial Light Modulator due to optical clipping inside the pulse shaper. Consequently, we developed a new set of orthogonal modes within the accessible spectral range, known as flat modes. These flat modes are orthogonal by design, and their size is determined by minimizing the difference with the theoretical HG modes. Surprisingly, they exhibited higher squeezing and anti-squeezing values.

%%%%%%%%%%%%We therefore expect that the measured squeezing values in Fig.~\ref{fig:SqzCurves} constitute a lower bound for the potential squeezing that can be achieved  with larger mode-matching visibilities and without optical clipping. This is also witnessed by the significant asymmetry in the squeezing and antisqueezing values measured in the flat mode basis.

This measurement demonstrates the experimental realization of optical multimode squeezed states composed of at least 21 HG frequency modes, as well as larger squeezing values in the set of defined flat modes.

\begin{figure}
    \includegraphics[width=\columnwidth]{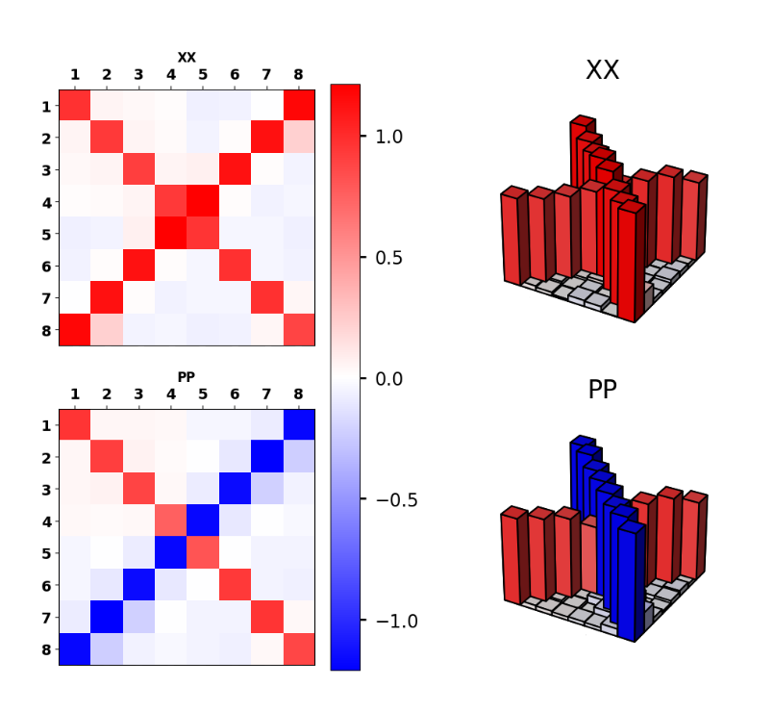}
    \caption{Expected values of the $\{\hat{x}_i\hat{x}_j\}$ and $\{\hat{p}_i\hat{p}_j\}$ quadrature components composing the covariance matrix of the multimode state, measured in the frexel basis using 8 frequency bands.}
    \label{fig:cov_mat}
\end{figure}
%\subsection{Covariance Matrix measurement}
In order to understand better the nature of the correlations in the quantum states, we studied the multimode field in the so-called \textit{frexel} basis, which is composed of 8 equally spaced (about 7 nm) frequency bands covering the total spectrum of the LO. Details about the measurement technique used can be found in \cite{Koadou23}. 

We recovered the covariance matrix of the state from the correlation measurement between the frexel quadrature operator components $\{\hat{x}_i\}$ and $\{\hat{p}_i\}$, where $i$ runs over the number of frexels. The correlation measurements of the type $\{\hat{x}_i\hat{x}_j\}$, $\{\hat{p}_i\hat{p}_j\}$  are shown in Fig.~\ref{fig:cov_mat}, while correlations of the type $\{\hat{x}_i\hat{p}_j\}$ are expected to be zero by symmetry arguments \cite{Patera10,Arzani18}.

The off-diagonal elements of the covariance matrix indicate the presence of correlations between the frequency bands. In order to test for quantum correlations, i.e. entanglement, we used a positive partial transpose (PPT) criterion \cite{Simon2000}. We obtained a violation of the PPT criterion, hence indicating entanglement, for 94\% of the possible bipartitions of the state.

A numerical diagonalization of the measured covariance matrix allows us to recover the experimental multimode squeezed basis and the squeezing distribution over the modes in this basis (also called \textit{supermodes} \cite{Roslund14}). The numerical eigenmodes are in good agreement with the HG modes expected from our theoretical analysis, with a slightly wider spectral width.

Information about the PPT criterion and the diagonalized eigenmodes and eigenvalues can be found in the Supplemental Material \cite{SM}.

%On the right of Fig.~\ref{fig:cov_mat}, we perform a numerical diagonalization of the measured covariance matrix to recover the eigenmode basis, where no entanglement is present. Thus we expect these eigenvectors to resemble discretized versions of the supermodes, with eigenvalues related to a squeezing level value over the frequency band composing the frexel.

%The eigenmodes and eigenvalues from the diagonalization of the covariance matrix  can be found in Appendix G.

\begin{figure}
    \centering
    \includegraphics[width=\columnwidth]{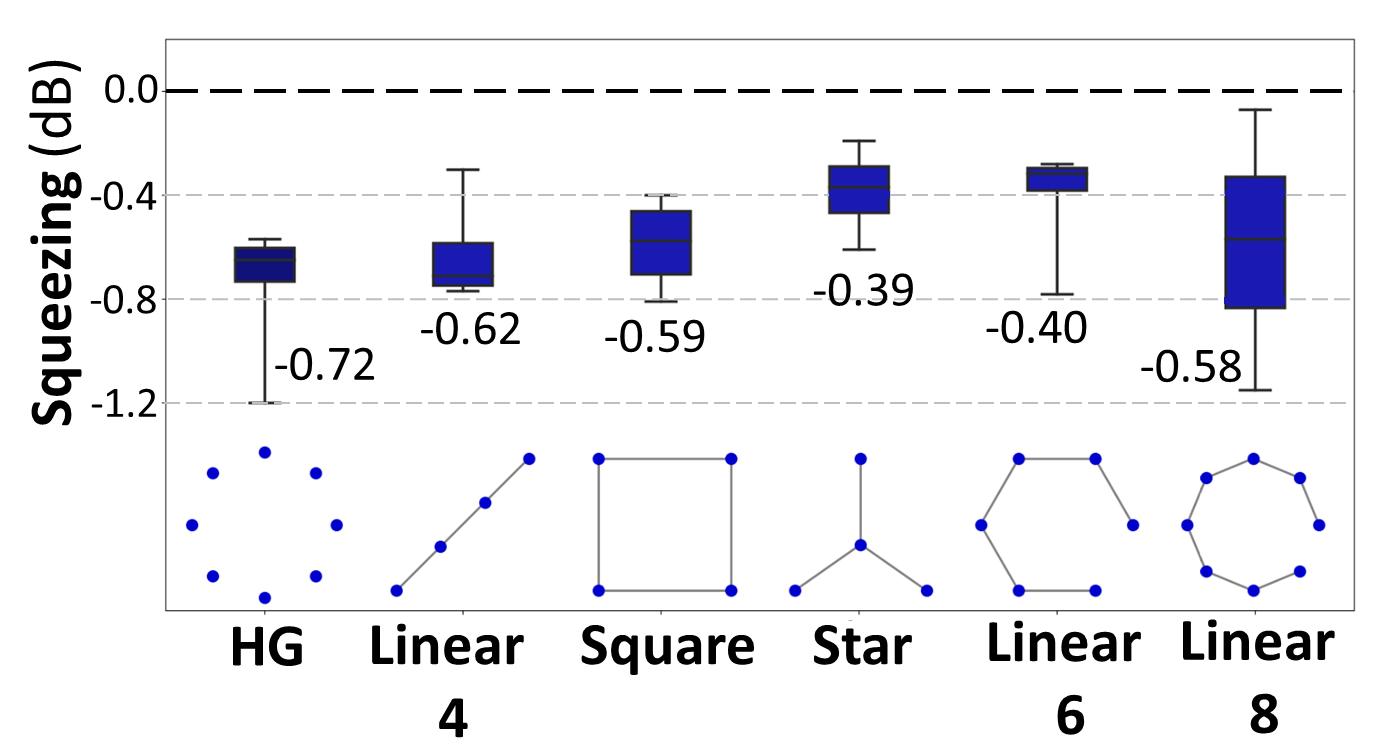}
    \caption{The boxplots sum up the squeezing value for the first HG modes (left), and the nullifiers of different cluster state size and topologies. The numerical values indicate the mean nullifiers's squeezing value, the black line spread over the squeezing values of the different nullifiers while the blue boxes are related to their variance (they give the amount of values in the interval between the first and third quartiles for each nullifier dataset). }
    \label{fig:cluster}
\end{figure}

As a note, the frexel basis is of interest, not only because it is easily accessible  via the  shaping capability of the LO,  but also because the frexel modes can be easily spatially separated via dispersive elements and sent to different locations, which is a requisite for potential multiparty quantum communication protocols.

%\section{Cluster State Generation}

Finally, we used the experimental setup for the deterministic generation of some few-node cluster states, as a proof of principle on the versatility of the source for quantum network generation.

We prove the generation of the cluster states by measuring squeezing in the nullifiers that characterize a specific adjacency matrix, \textit{i.e.}, a particular topology defining the graph. Changing from one topology to another can be achieved by appropriately changing the mask on the pulse shaper  \cite{vanLoock2007}. The nullifiers, $\{\hat{\delta}_i\}$, of a particular graph with quadratures operators for the nodes denoted $\hat{x}_i$ and $\hat{p}_i$ are written as:
\begin{equation}
    \hat{\delta}_i = \hat{p}_i - \sum_j V_{ij}\hat{x}_j,
\end{equation}
where $V_{ij}$ are the elements of the adjacency matrix defining the cluster state, which is known in advance \cite{Menicucci11}. Furthermore, the cluster state quality can be qualitatively tested by the amount of squeezing measured in the nullifiers.

We measured 4-node cluster states with different topologies and whose nullifier's level of squeezing is summarized in Fig.~\ref{fig:cluster} via statistical box plots for each cluster topology. Additionally, we show that all nullifiers are squeezed below the shot noise up to the 8-mode linear cluster. The squeezing levels of the nullifiers are smaller than the levels presented in Fig.\ref{fig:SqzCurves}. We suspect this is due to the fact that the used HG modes present some residual correlations, hence affecting the quality of the clusters that are built from that basis. We nevertheless leave further source optimization for future studies and protocol applications. Details about the measurement of the cluster states can be found in the Supplemental Material \cite{SM}.
 
%In Fig.(\ref{fig:cluster}), we effectively see squeezing in the nullifiers for five different cluster topologies and hence proving experimentally the deterministic generation of photonic cluster states in the frequency domain.

%\section{Conclusion}
In conclusion, we report on a deterministic source of multimode quantum states of light generated via type-0 SPDC in a nonlinear ppKTP rectangular waveguide. We generated a multimode squeezed vacuum state from which we measured up to 21 squeezed optical spectral  modes in the HG basis and over 2 dB of squeezing on four spectral \textit{flat} modes. We measured and diagonalized the covariance matrix in the frexel basis and demonstrated the realization of different cluster states topologies, characterizing them with the squeezing in their nullifier operators.

The quality of the quantum networks can be further enhanced by improving the signal-LO mode-matching and the optical configuration of the pulse shaper, both within reach of current technology. 

The demonstrated source paves the way towards the implementation of multiparty, entanglement-based quantum communication protocols as well as building scalable resources for quantum computing at telecom wavelenghts.  

%\begin{acknowledgments}
This   work   was   supported   by   the   European   Research Council under the Consolidator Grant COQCOoN (Grant No.  820079).
%\end{acknowledgments}
\bibliography{mybibliography}
%\begin{comment}
\clearpage

\appendix

\section{Numerical simulation of the multimode state in our experimental configuration}

Numerical simulations were performed in order to predict the properties of the independent squeezed modes at the output of the waveguide.

The quantum state after a SPDC interaction is determined by the joint-spectral amplitude (JSA):
\begin{equation}
    J(\omega_s,\omega_i) = \sum_{k}\lambda_{k} h_{k}(\omega_s)g_{k}(\omega_i),
    \label{Schm_decom}
\end{equation}
where $\omega_{s/i}$ is the signal/idler frequency, $\lambda_k$ are the Schmidt coefficients and $h_k(\omega_s)$ ($g_k(\omega_i)$) are the signal (idler) frequency modes composing the signal (idler) field after the interaction. In the degenerate case, as it is the case in this work, signal and idler fields are identical and there is a single outfield field after the interaction.

\begin{figure*}
	\begin{center}
		\includegraphics[width=0.9\textwidth]{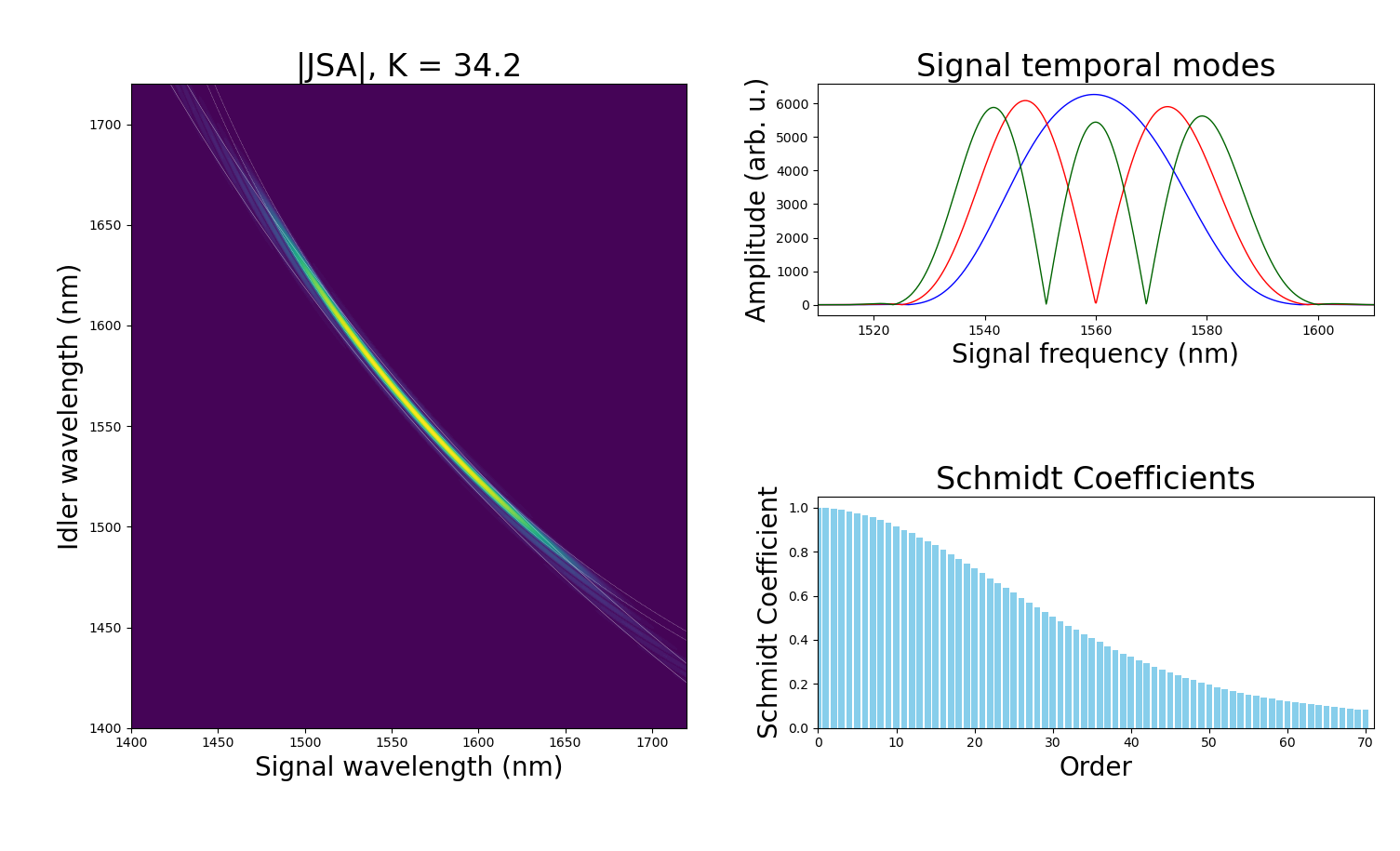}
	\end{center}
	\caption{\textbf{Numerical simulation of the independent squeezed states}. Left: JSA function, Right: First three frequency modes, approximating Hermite-Gauss modes, and the distribution of the Schmidt coefficients.}\label{fig:JSA}
\end{figure*}

Given the experimental values measured from the second harmonic field (hence the pump to the ppKTP waveguide), and the chosen waveguide dimensions (3 by 3 $\mu$m and 15 mm in length), 
%a numerical simulation was performed, whose 
the results of the numerical JSA and its singular value decomposition are shown in Fig.~\ref{fig:JSA}. A number of about 34 modes is expected with the Schmidt distribution, $\{\lambda_k\}$, shown in the figure. The first three frequency modes, approximating Hermite-Gauss modes, are also shown. For more details about the numerical simulation of the nonlinear waveguides and how the experiment was designed, see~\cite{VRR21}.

\section{Waveguide Characterization}

\begin{figure*}
\centering
		\includegraphics[width=0.8\textwidth]{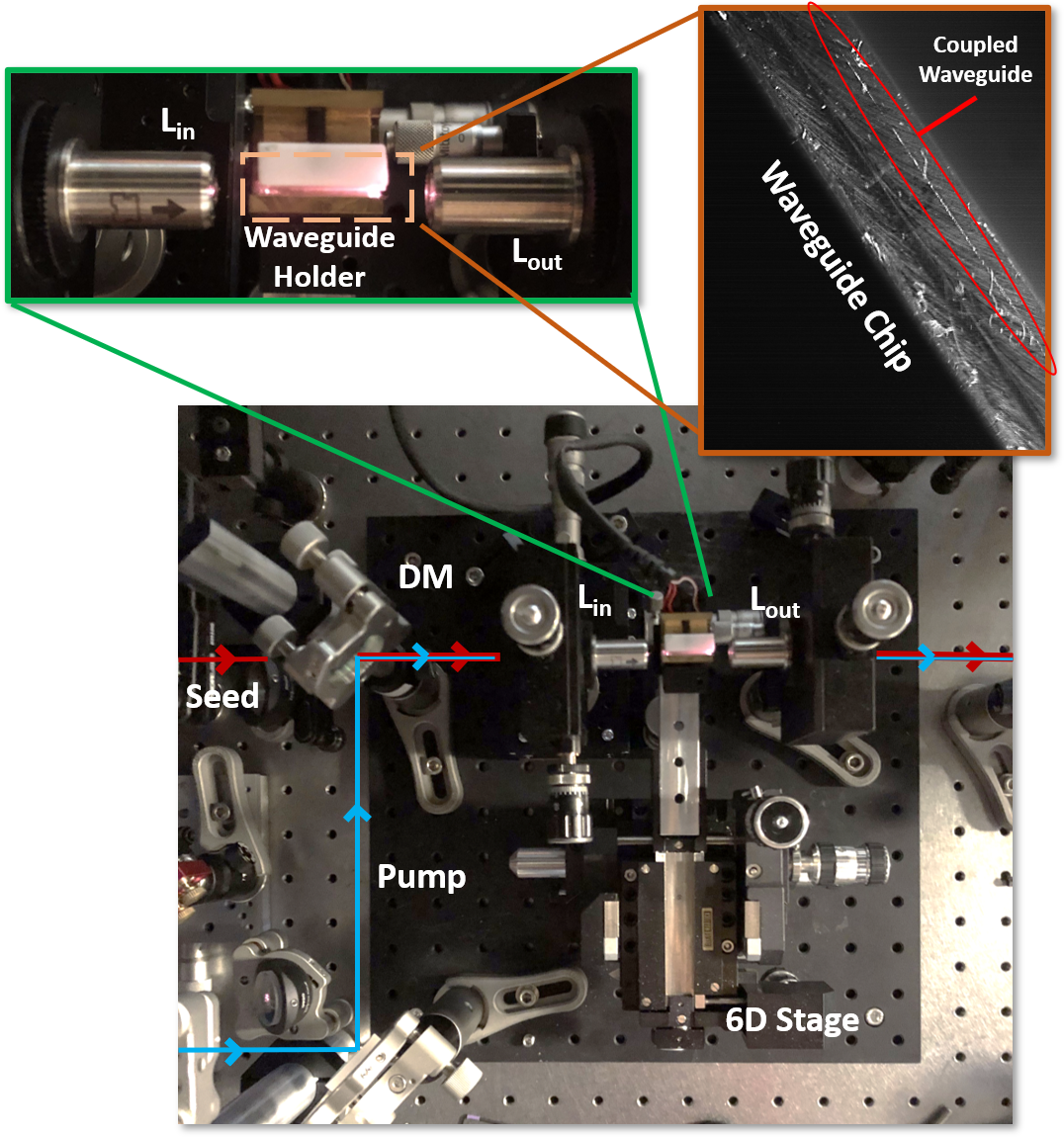}
	\caption{\textbf{Waveguide characterization setup.} The seed (1560 nm) and the pump (780 nm) are coupled to the rectangular waveguide showed in the zoom at the upper right corner via the two lenses $L_{in}$ and $L_{out}$. The waveguides are mounted on a 6D stage that allows sub-micrometric alignment for the waveguide couplings to both seed and pump fields. DM stands for Dichroic Mirror.}\label{fig:waveguide_setup}
\end{figure*}

\begin{figure}
\centering
		\includegraphics[width=\columnwidth]{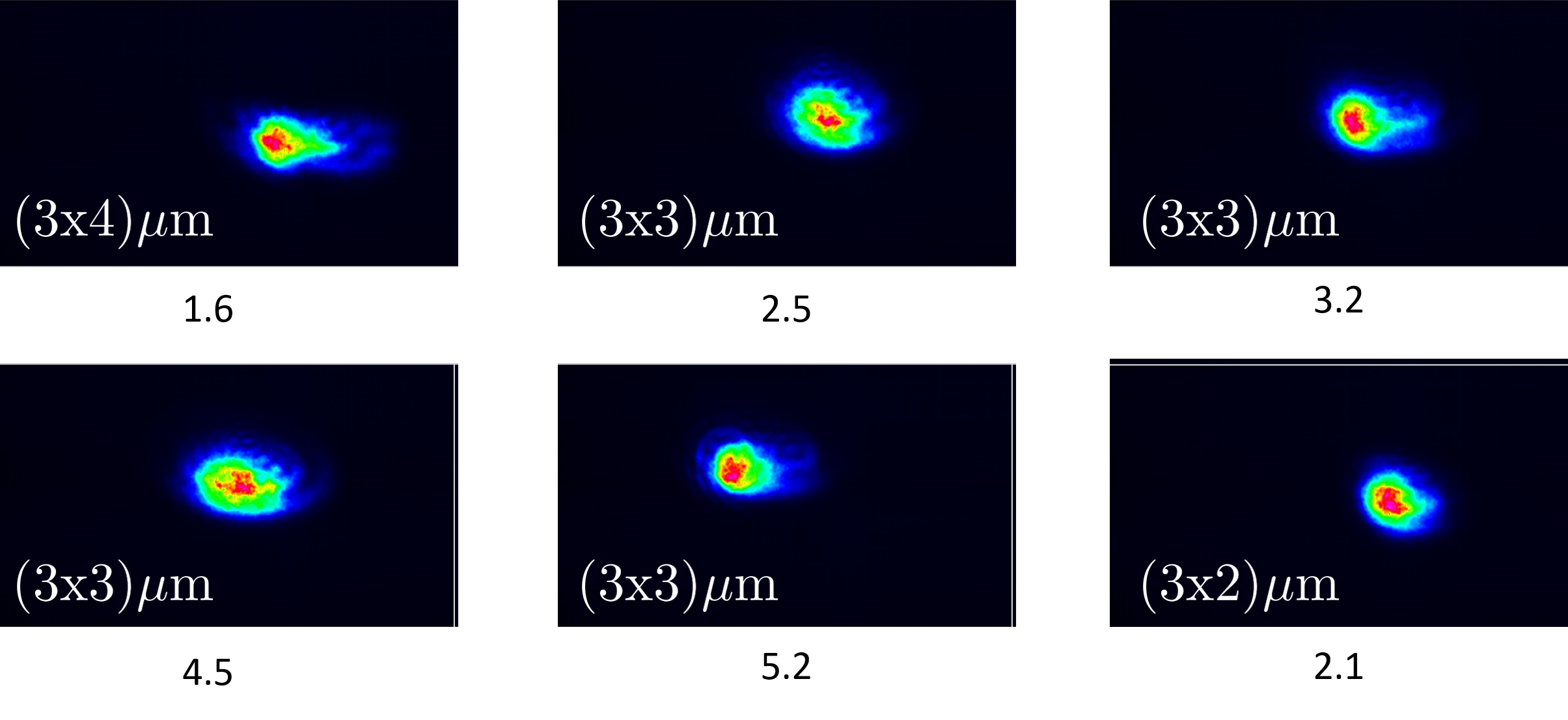}
	\caption{\textbf{Spatial modes at 1560 nm.} Examples of the spatial modes measured after the waveguide coupling at telecom wavelengths for 6 different waveguides.}\label{fig:spatial_telecom}
\end{figure}

The characterization of our waveguides was performed by coupling them to the C-band wavelength (central wavelength: 1560 nm), as well as the second harmonic field (central wavelength: 780 nm). See Fig.\ref{fig:waveguide_setup} for a physical photo of the waveguide setup in the lab. A measurement of the spectrum of the second harmonic field produced by the C-band input gives a qualitative estimate on the homogeneity of the waveguide along the propagation direction (a sinc-like function is to be measured, which is the response to a square nonlinear profile by the Fourier transform, expected from a homogeneuous non-linear coefficient). The amount of second harmonic produced from the telecom input gives an estimate on the nonlinear coefficient of the waveguide, that is fitted from data in the next section. Finally, a measurement of the spatial profile of the output telecom light (see Fig.\ref{fig:spatial_telecom} for some examples of this measurement) can be compared with a numerical simulation (in our case with a finite element method implemented based on \cite{Fallahkhair08}) to check the single-spatial-mode feature of the waveguides at telecom wavelengths, and the deviation from the expected fundamental mode propagating in the actual structure.

The homodyne measurement was performed with a home-made detector, including the two photodetectors and the transimpedance circuit outputting the homodyne electrical signal. The total homodyne efficiency, $\eta_{\mathrm{h}}$, can be decomposed in the following terms:
\begin{equation}
    \eta_{\mathrm{h}} = \eta_{\mathrm{PD}}\cdot\eta_{\mathrm{el}}\cdot\eta_{\mathrm{opt}}\cdot\eta_{\mathrm{mod}},
    \label{eq:hom_eff}
\end{equation}
where $\eta_{\mathrm{PD}}$ is the photodetector's efficiency (around 85\% in our case), $\eta_{\mathrm{opt}}$ is related to the optical losses in the homodyne circuit (near unity in our case), $\eta_{\mathrm{mod}}$ is the mode matching efficiency, \textit{i.e.}, how close are the LO and the signal in terms of polarization, spatial and temporal profile when interfering, and $\eta_{\mathrm{el}}$ is the electronic efficiency.

The electronic efficiency can in turn be written as $\eta_{\mathrm{el}}=1-1/\mathrm{SNR}$ \cite{Kumar12}, with $\mathrm{SNR}$ the signal to noise ratio, \textit{i.e.}, the ratio between the shot noise at a certain input intensity and the value in the absence of any input signal (electronic noise). We measured the best signal-to-noise ratio (also called clearance, if one measures it in dB) at a demodulation frequency of 2 MHz, where the clearance was about 20 dB at 2 mW of input power.

The mode-matching efficiency was limiting the amount of available squeezing in our experiment as it is discussed in the main text. Such term scales quadratically with the visibility between LO and signal and it can be measured by mimicking the signal via an intense beam following the same path as the signal in the experiment, and making it interfere with the LO. The $77 \%$ value reported in the main text has be measured with a beam matching the spectral width of the LO. In the case of the measurement of the Hermite-Gauss basis, for higher-order modes we expect a lower mode-matching efficiency, since this spectral matching is poorer.

\section{Phase Sensitive Amplification Results}

Before measuring the multimode squeezing curves shown in the main text, we built a degenerate Optical Parametric Amplifier, or OPA, by pumping the waveguide with a relatively intense seed field at telecom wavelengths (taken directly from the ultrafast laser) and a pump field (from our second harmonic ppLN crystal at 780 nm).

\begin{figure}
\centering
		\includegraphics[width=\columnwidth]{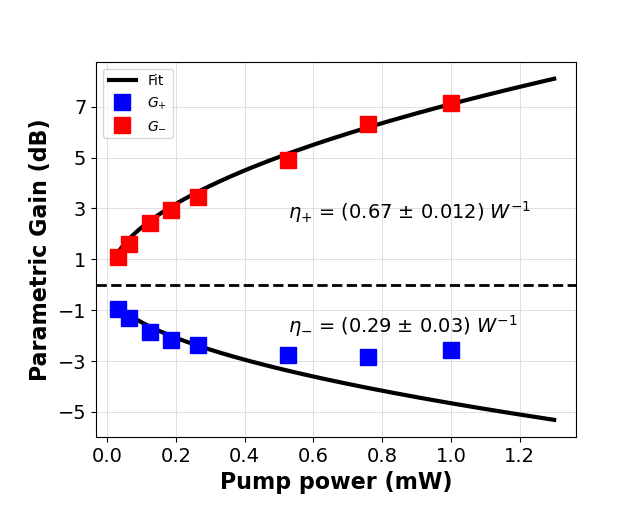}
	\caption{Parametric gain vs pump power. Extrema in the modulation of the amplitude for the degenerate OPA, fit to Eq.~(\ref{G+-}) gives the parametric efficiency, related to the total squeezing level.}\label{fig:ParametricGain}
\end{figure}

The phenomenon of parametric amplification can be observed as a modulation of the seed amplitude at the output of the waveguide, depending on the relative phase between the seed and the pump fields. This measurement allows us to show the existence of parametric gain in our waveguides, which is a precondition of squeezing generation, even though the levels of multimode squeezing cannot be predicted in this way.

The extrema of the parametric gain, $G_{\pm}$, can be approximated, for a single-mode OPA, as \cite{G+-,Umeki11}:
\begin{equation}
    G_{\pm} \sim \exp(\pm2\sqrt{\eta_{\mathrm{PSA}}P})
    \label{G+-},
\end{equation}
with $P$ the pump power and $\eta_{\mathrm{PSA}}$ the parametric efficiency. Ideally, the minimum deamplification, $G_{-}$, should be symmetric with respect to $G_{+}$, although a disparity between the two has been attributed to a distortion of the spatial or temporal profile inside the nonlinear material in previous experiments. The disparity appears at sufficiently high power density in the material.

Fig.~\ref{fig:ParametricGain} shows the parametric gain measured as a function of the pump power. The data fits well Eq.~(\ref{G+-}) and gives two values for the parametric efficiency due to their asymmetry. A measurement of the second harmonic efficiency in the same nonlinear waveguide gives an efficiency of $\eta_{\mathrm{SHG}}=0.33$ W$^{-1}$, which is in good agreement with the extracted parametric efficiency for the deamplification.

The conclusion of the phase sensitive experiment is that, at pump powers of few mW, we can expect at least some dB of total squeezing in our multimode states, which are the final measured values showed in the main text.

\section{Squeezing curves}

\begin{figure}
    \centering
    \includegraphics[width=\columnwidth]{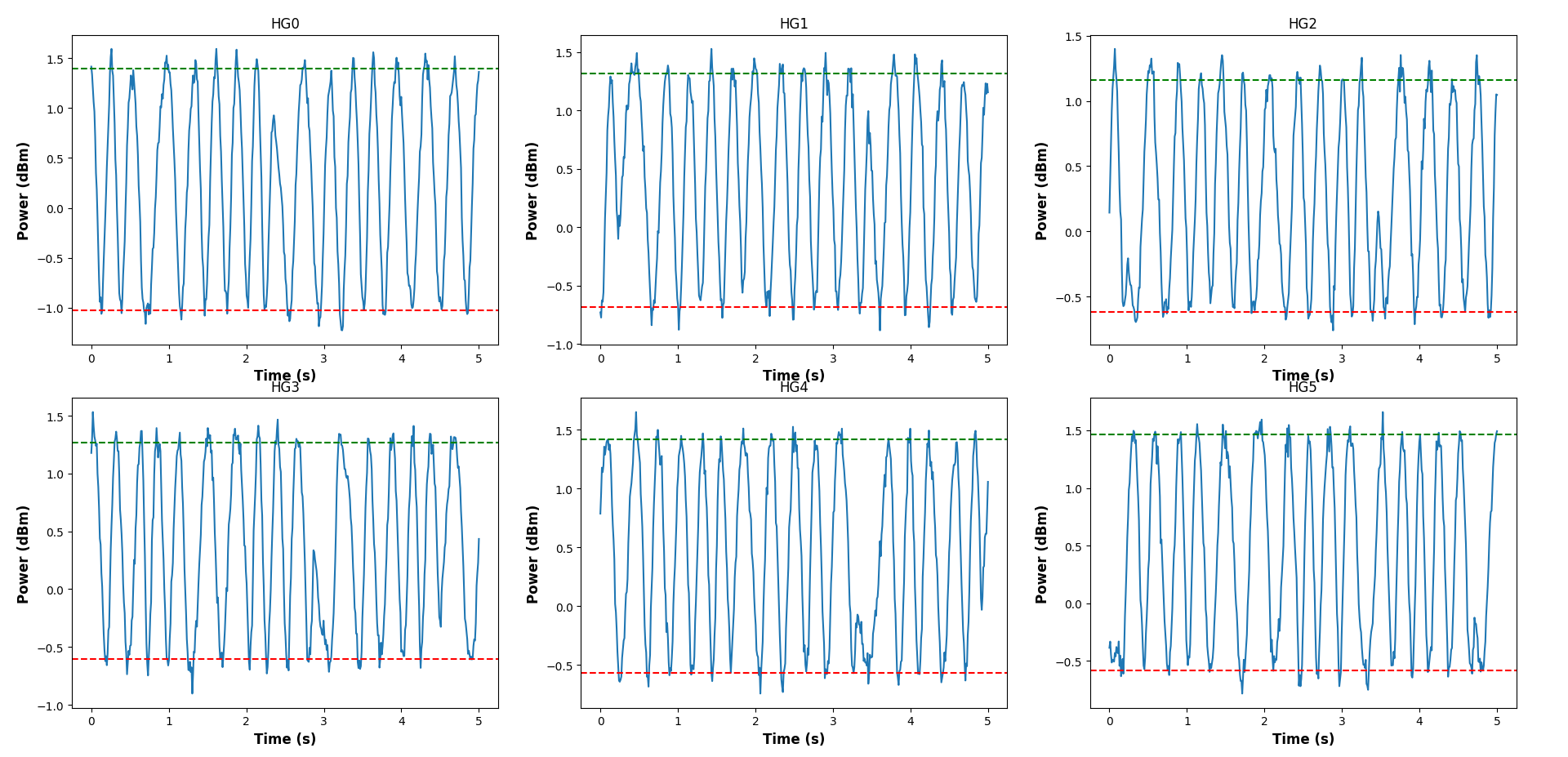}
    \caption{Squeezing traces for the projection onto the first six Hermite Gauss modes. The average squeezing and anti-squeezing values are shown as the dotted red and green horizontal lines.}
    \label{fig:sqz_curves}
\end{figure}

The squeezing (antisqueezing) values presented in the main text were obtained by averaging 15 minima (maxima) from the modulated phase signal for each Hermite-Gauss mode, performing homodyne detection with a linear phase shift in time via a piezoelectric mirror. Examples of such traces are depicted in Fig.~\ref{fig:sqz_curves} for the first few modes. The mean squeezing and anti-squeezing values where computed after applying a Savitsky-Golay filter to smooth the data without distorting the signal avoiding unwanted local extrema.The local extrema generated by the piezoelectric mirror when changing direction were intentionally disregarded.

\section{Independent Hermite-Gauss modes}

An experimental detrimental effect in the setup described in this work is what we call \textit{optical clipping}, which arises due to the limited size of a cylindrical mirror responsible for focusing the light beam onto the SLM for pulse shaping.  Since the beam comes from a refractive element (a grating), some wavelengths at the extremes of the spectrum are cut off. Due to this effect, the LO spectral extremes cannot be used in homodyne detection (the corresponding clipped regions from the different modes are shown in red in Fig.~\ref{fig:eigenmodes}). We would then expect even larger values for the measured squeezing levels if the full spectrum was experimentally available.

As a consequence of optical clipping, the 21 Hermite-Gauss modes implemented in the experiment were not exactly orthogonal, with the effect being more pronounced the larger the mode order. Therefore, the actual number of measured modes is expected to be lower than 21. This number depends on the dimensionality of the vector space spanned by the non-orthogonal modes. In order to account for that, we evaluate the rank of the matrix composed by these modes performing a singular value decomposition on the set of 21 vectors after measuring the non-available interval of wavelengths and subtracting it from the HG spectrum. The non-zero singular values count the number of linearly independent modes and hence the dimension of the space we are looking for. The singular values resulting from the decomposition is summarized in Table \ref{svd_table}.

%\begin{table*}
  %\centering
  %\caption{Singular values obtained from clipped Hermite-Gauss modes used in the experiment.}
  %\begin{tabular}{|*{12}{c}|}
    %\hline
     %Singular Value & 0 & 1 & 2 & 3 & 4 & 5 & 6 & 7 & 8 & 9 & 10 \\ [0.5 ex] 
    %\hline
    %& 1.42\ \ & 1.40\ \  & 1.31\ \  & 1.28\ \  & 1.21\ \  & 1.19\ \  & 1.18\ \  & 1.12\ \  & 1.09\ \  & 1.08\ \  & 1.02\ \ \\
    %\hline
    %Singular Value & 11 & 12 & 13 & 14 & 15 & 16 & 17 & 18 & 19 & 20 \\ [0.5 ex] 
    %\hline
    %& 0.99\ \  & 0.97\ \  & 0.94\ \  & 0.90\ \  & 0.84\ \  & 0.60\ \  & 0.30\ \  & 0.08\ \  & 0.02\ \  & 0.005\ \  \\
    %\hline
  %\end{tabular}
  %\label{svd}
%\end{table*}
\begin{table}
\begin{tabular}{| c | c |} 
 \hline
 \textbf{ Mode Number} & \textbf{ Singular Value} \\ [0.5ex] 
 \hline\hline
0 & 1.42  \\ 
 \hline
 1 & 1.40  \\ 
 \hline
 2 & 1.31  \\
 \hline
 3 & 1.28  \\
 \hline
 4 & 1.21  \\
 \hline
  5 & 1.19  \\
 \hline
  6 & 1.18  \\
 \hline
  7 & 1.12  \\
 \hline
  8 & 1.09  \\
   \hline
 9 & 1.08  \\
 \hline
  10 & 1.02  \\
 \hline
  11 & 0.99  \\
 \hline
  12 & 0.97  \\
 \hline
  13 & 0.94  \\
   \hline
 14 & 0.90  \\
 \hline
  15 & 0.84  \\
 \hline
  16 & 0.60  \\
 \hline
  17 & 0.30  \\
 \hline
  18 & 0.08  \\
   \hline
  19 & 0.02  \\
 \hline
 20 & 0.005 \\ \hline
\end{tabular}
\caption{Singular values for the first 21 Hermite-Gauss modes after optical clipping.}
\label{svd_table}
\end{table}

Since this is a numerical computation, we need a somewhat arbitrary criterion to give a whole number indicating the dimension of our vector space from this result. In our case, we obtain the dimension of the vector space by accounting for singular values that are at least 10\% of the highest one. This criterion gives us 18 linearly independent modes. The value being close to the 21 Hermite-Gauss modes indicates that the optical clipping did not have a large impact in reducing the dimensionality of the modes measured in the experiment.

Although the optical clipping is not a fundamental limitation of the setup (it can be solved by the use of a larger cylindrical mirror), it did limit the maximum number of orthogonal modes that we could confidently measure in the experiment. That is the reason why we stop showing squeezing for mode orders larger than 21.

\section{Flat mode basis}

\begin{figure}
    \centering
    \includegraphics[width=\columnwidth]{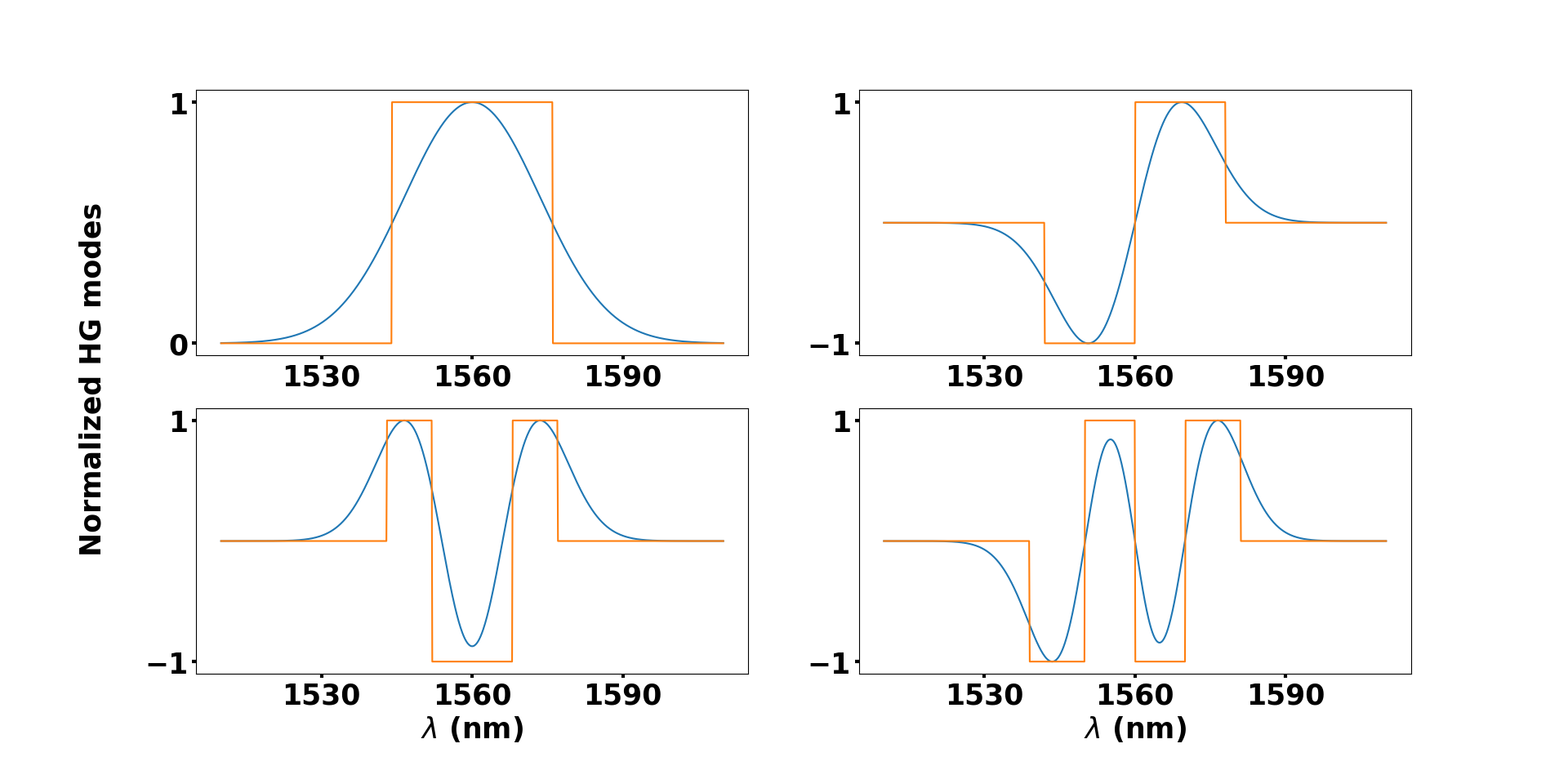}
    \caption{First 4 Hermite-Gauss modes measured (blue) and the corresponding spectral flat modes obtained from them (orange).}
    \label{fig:flat}
\end{figure}

The spectral flat mode basis discussed in the main text is constructed to be orthogonal, taking advantage of even and odd symmetries for the subsequent mode functions. It was specifically designed to roughly resemble the first 4 Hermite-Gauss modes, so that the spectral width of each flat mode was obtained by minimizing the $l^2$ norm distance to the corresponding Hermite-Gauss mode counterpart. Fig.~\ref{fig:flat} shows the 4 flat modes constructed and implemented in the experiment, together with the first 4 Hermite-Gauss modes.

\section{Eigenmodes from the diagonalization of the covariance matrix}

The numerical diagonalization of the covariance matrix performed in the main text gives back the eigenmode basis where there are no quantum correlations, sometimes called the supermode basis. For additional information concerning the reconstruction of the covariance matrix see \cite{Koadou23}.

We therefore expect the numerical eigenmodes to resemble discretized versions of the quasi-Hermite-Gauss supermodes found in our numerical simulations.

\begin{figure}
    \centering
    \includegraphics[width=\columnwidth]{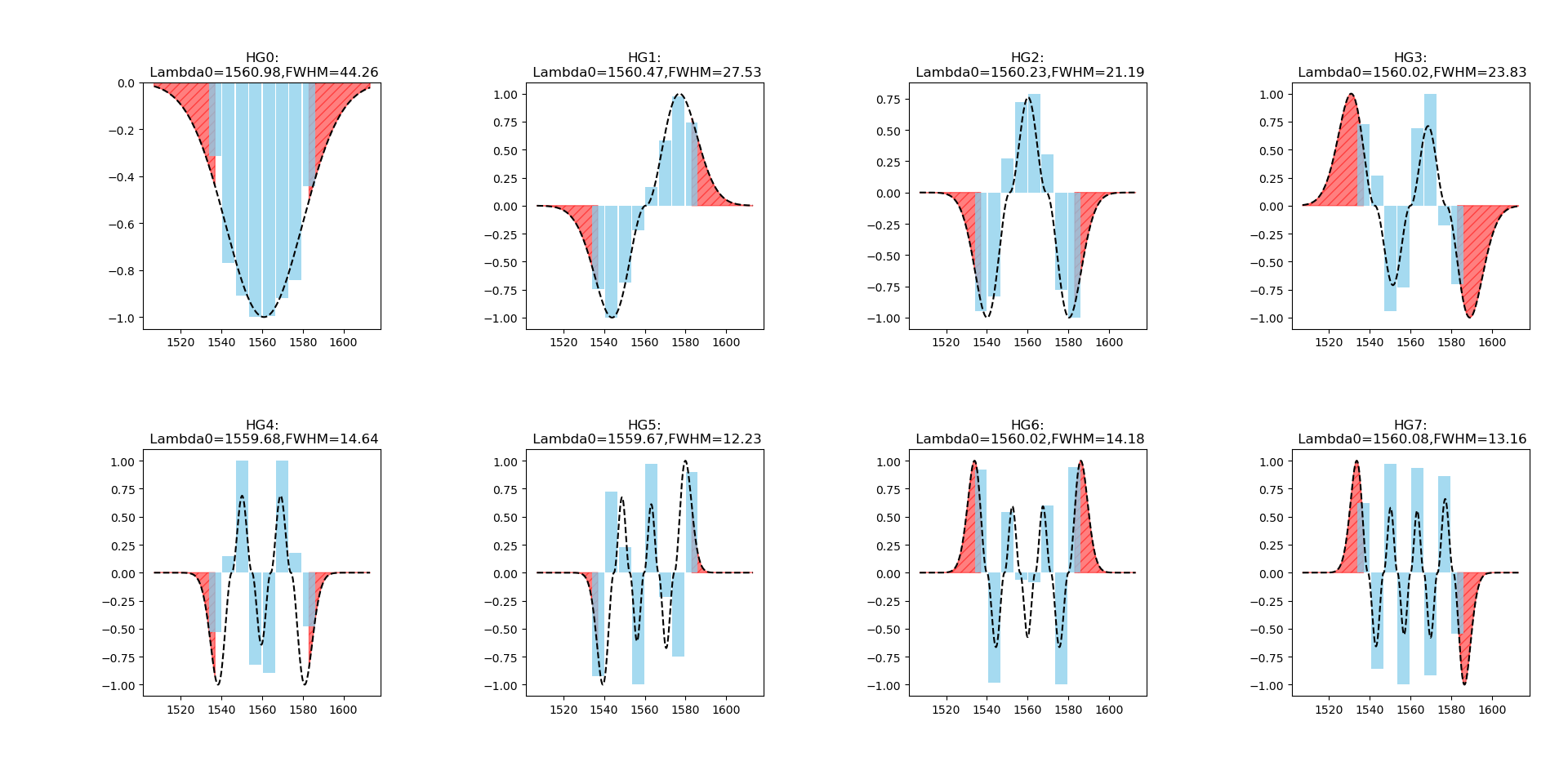}
    \caption{Eigenmodes of the covariance matrix and a fit to HG modes. The wavelength outside of the Local Oscillator bandwidth due to optical clipping is depicted as the red areas.}
    \label{fig:eigenmodes}
\end{figure}

 Fig.~\ref{fig:eigenmodes} shows the numerical eigenmodes obtained after the diagonalization. In general, the supermode shapes are in very good agreement with the expected theoretical Hermite Gauss modes. The bandwidth of every Hermite-Gauss is systematically higher than the theoretical value and it's slightly higher than the total LO bandwidth for higher order modes, which can detriment the real value of squeezing present in the state compared to the presented measured values.

 %On a technical note, the 55 nm bandwidth of our LO was slightly smaller than the bandwidth of our expected quantum signal. This limits the experimentally accessible modes if the wavelengths outside the LO bandwidth are involved in the spectral features of the supermode. This effect is noticeable for high-order modes, which are the most broadband. In Fig.~\ref{fig:eigenmodes}, the red area highlights  the wavelength range that was not accessible with our LO bandwidth.
 
\begin{figure}
\centering
   \includegraphics[width=\columnwidth]{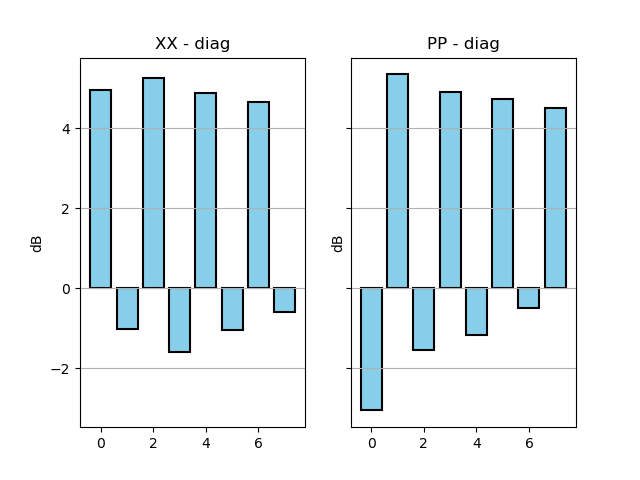}
   \caption{Squeezing values obtained from the numerical diagonalization of the covariance matrix in the frequency basis. The modes have been arranged in decreasing order of antisqueezed quadratures, and the first eigenmode is squeezed in the $\hat{p}$ quadrature by convention.}
    \label{fig:sqz_numerical}
\end{figure}

Note that the eigenvalues of the covariance matrix are directly related to the expected squeezing levels of each eigenmode. Fig.~\ref{fig:sqz_numerical} shows the eigenvalues obtained in the diagonalization, which are consistent with those values directly measured in the HG basis from the main text.

\section{Cluster state measurement}

As stated in the main text, we measured some few-nodes cluster states with different topologies corresponding to different adjacency matrices. The specific adjacency matrices used in this work are the following:

%Adjancency matrices
\begin{equation}
    V_{\text{linear 4}} = \begin{pmatrix}
    0 & 1 & 0 & 0 \\
    1 & 0 & 1 & 0 \\
    0 & 1 & 0 & 1 \\
    0 & 0 & 1 & 0
\end{pmatrix}
\end{equation}
 
\begin{equation}
V_{\text{square}} = \begin{pmatrix}
    0 & 1 & 0 & 1 \\
    1 & 0 & 1 & 0 \\
    0 & 1 & 0 & 1 \\
    1 & 0 & 1 & 0
\end{pmatrix}
\end{equation}
\begin{equation}
V_{\text{star}} = \begin{pmatrix}
    0 & 1 & 1 & 1 \\
    1 & 0 & 0 & 0 \\
    1 & 0 & 0 & 0 \\
    1 & 0 & 0 & 0
\end{pmatrix} 
\end{equation}

\begin{equation}
V_{\text{linear 6}} = \begin{pmatrix}
    0 & 1 & 0 & 0 & 0 & 0 \\
    1 & 0 & 1 & 0 & 0 & 0 \\
    0 & 1 & 0 & 1 & 0 & 0 \\
    0 & 0 & 1 & 0 & 1 & 0 \\
    0 & 0 & 0 & 1 & 0 & 1 \\
    0 & 0 & 0 & 0 & 1 & 0
\end{pmatrix}
\end{equation}
\begin{equation}   
V_{\text{linear 8}} = \begin{pmatrix}
    0 & 1 & 0 & 0 & 0 & 0 & 0 & 0 \\
    1 & 0 & 1 & 0 & 0 & 0 & 0 & 0 \\
    0 & 1 & 0 & 1 & 0 & 0 & 0 & 0 \\
    0 & 0 & 1 & 0 & 1 & 0 & 0 & 0 \\
    0 & 0 & 0 & 1 & 0 & 1 & 0 & 0 \\
    0 & 0 & 0 & 0 & 1 & 0 & 1 & 0 \\
    0 & 0 & 0 & 0 & 0 & 1 & 0 & 1 \\
    0 & 0 & 0 & 0 & 0 & 0 & 1 & 0 \\
\end{pmatrix}    
\end{equation}

We construct the nullifiers starting from the Hermite-Gauss basis used in the main text. The modes represented by the nodes of the cluster state are then a linear superposition of the Hermite-Gauss modes.  The bandwidth of the nodes in the cluster is therefore the same as the presented HG bandwidth. The nullifiers are combination of quadrature operators of the node of the cluster, so that the measurement of their variance is done by setting as the LO mode a specific combination of the modes corresponding to the node of the clusters.  We then measured the squeezing of each nullifier of a particular cluster state topology, i.e. a particular adjacency matrix, by averaging 15 squeezing traces, in the same manner as when measuring squeezing in the HG modes.  The squeezing values of the nullifiers of a given cluster are expected to be very close, in particular they are expected to be the same for a perfect cluster. Therefore in the main text we use the statistical box plot representation for the mean squeezing value and its variance over all the nullifiers in the cluster.

As a technical note, we chose unweighted and undirected graphs in this text, thus assuming equal interaction strength between all nodes for the generated clusters. 

\section{Peres–Horodecki (PPT) criterion}

The Positive Partial Transposition (PPT) criterion can be checked from the covariance matrix measurement to determine whether quantum correlations are present in our state in the frexel basis.

The PPT criterion is based on the fact that given a density matrix defining a quantum state, the partial transpose matrix of two possible bipartitions of the density matrix has to be positive defined for the two bipartitions being separable. The criterion applied to continuous variable systems can be found, for example, in \cite{Simon2000}. This theorem applies to the covariance matrix since it completely defines any Gaussian state.

In our case, given that we have an 8x8 matrix in our frexel basis, we have a total of 127 possible bipartitions. For each bipartition, we can define the minimum eigenvalue of the partial transpose matrix as the PPT value, indicating quantum correlations if negative. 

Fig.~\ref{fig:ppt_criterion} shows the PPT value for the different bipartitions of our covariance matrix, showing the violation of the PPT criterion, \textit{i.e.}, negativity in the partial transpose matrix, in 94\% of them.

More details about the PPT criterion applied to the covariance matrix can be found in \cite{Medeiros2014}.
\begin{figure}
\centering
    \includegraphics[width=1\columnwidth]{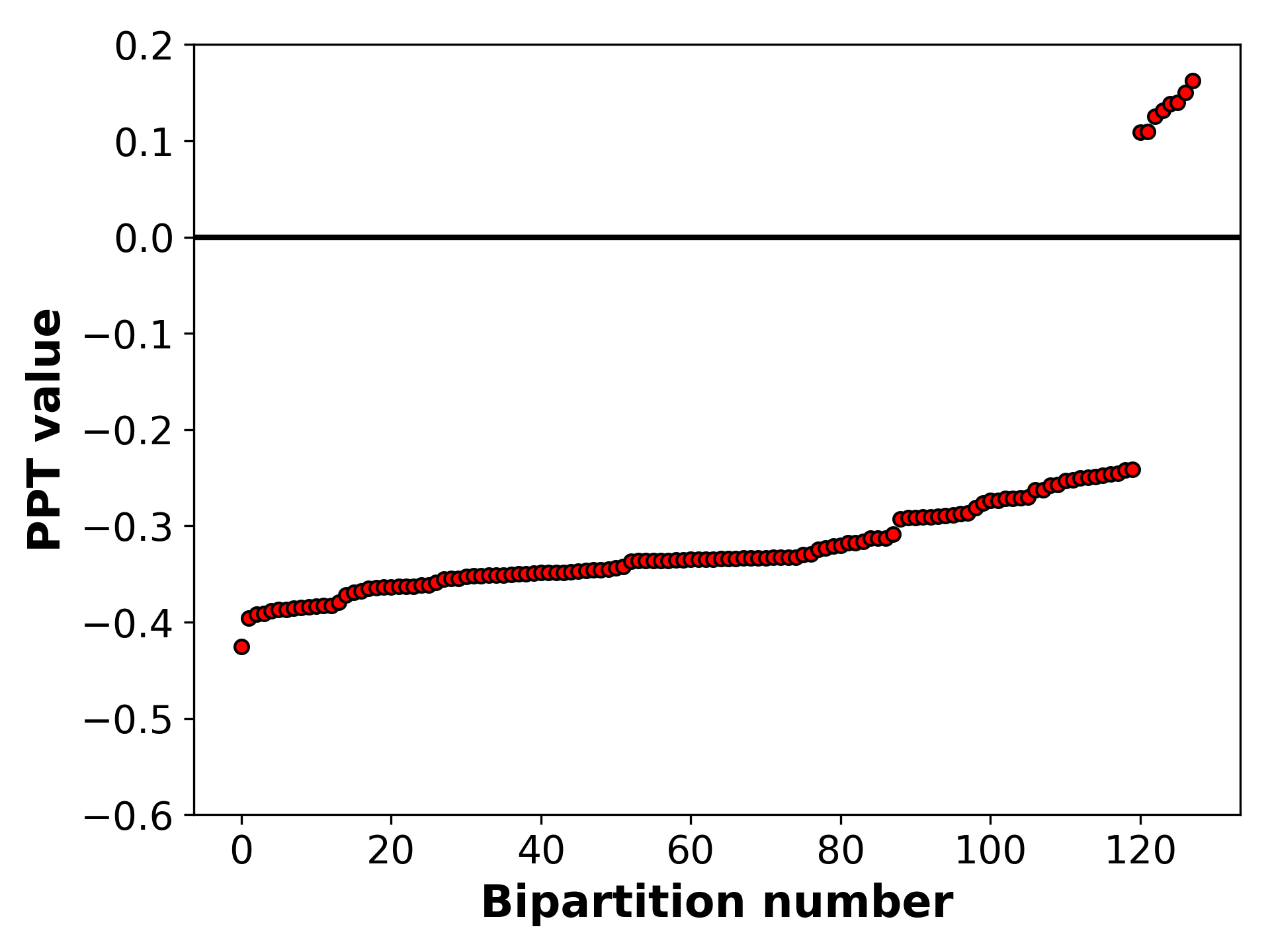}
    \caption{PPT values computed for all 127 possible bipartitions. The results show that almost all (119) of the values are negative, thereby indicating the non-separability of the state.}
    %\label{fig:ppt_criterion}
\end{figure}
%\end{comment}
\end{document}